\documentclass[12pt,a4paper]{iopart}
\usepackage{bm}
\usepackage{epsfig}

\usepackage{bm}
\usepackage{epsfig}
\usepackage{iopams}
\usepackage{amssymb}
\usepackage{setstack}

\usepackage{color}

\newcommand{\ee}{\mathrm{e}}

\newcommand{\CC}{\mathcal{C}}
\newcommand{\WW}{\mathbb{W}}

\newcommand{\mob}{m}

\begin{document}
\title 
{Geometrical interpretation of fluctuating hydrodynamics in diffusive systems} 
\author{Robert L. Jack}
\address{Department of Physics, University of Bath, Bath BA2 7AY, United Kingdom}
\author{Johannes Zimmer}
\address{Department of Mathematical Sciences, University of Bath, Bath BA2 7AY, United Kingdom}

\begin{abstract}
  We discuss geometric formulations of hydrodynamic limits in   diffusive systems.  Specifically, we describe a geometrical   construction in the space of density profiles --- the Wasserstein   geometry --- which allows the deterministic hydrodynamic evolution   of the systems to be related to steepest descent of the free energy,   and show how this formulation can be related to most probable paths   of mesoscopic dissipative systems.  The geometric viewpoint is also   linked to fluctuating hydrodynamics of these systems via a saddle   point argument.
\end{abstract}
\maketitle

\section{Introduction}

In many physical situations, one seeks to describe a complex system of many components by a simpler theory that operates on large length (and time) scales.  Familiar examples include the description of molecular liquids by the continuum equations of fluid dynamics, or the use of a diffusion equation to describe the spreading of particles through a system.  These are examples of hydrodynamic limits, where systems of discrete particles can be modelled by the evolution of continuous fields, such as local density and velocity.
Here, we concentrate on a family of microscopic models with overdamped (stochastic) dynamics, in which hydrodynamic limits have been studied over many years~\cite{spohn83,kov89,eyink1}: these models have also attracted considerable recent interest~\cite{bertini0102,bertini09,bertini14,TKL-prl07,   TKL,imparato09,lecomte10}.  In these \emph{diffusive systems}, the theory of ``fluctuating hydrodynamics''~\cite{spohn83} captures the behaviour in the hydrodynamic limit, including both deterministic and stochastic effects.  The models have been studied both in equilibrium and non-equilibrium settings. For example, they may be coupled to particle reservoirs with different chemical potentials, so that currents flow through the system.  Several elegant results have been derived, including the exact analysis of long-ranged correlations that appear in the non-equilibrium states~\cite{bertini0102,bertini09,bertini14,TKL-prl07,TKL}, and calculations of large deviations of currents and dynamical activities at equilibrium~\cite{imparato09,lecomte10}.

The aim of this article is to show how recent results from mathematics~\cite{jko,otto2001,ags} (see also~\cite{adams2011,adams2013}) provide a geometrical interpretation of the theory of fluctuating hydrodynamics for these systems.  In particular, we describe distance measures (metric structures) in the space of density profiles, such that the hydrodynamic limit equations for several diffusive systems correspond to ``steepest descent'' processes, or ``gradient flows''.  That is, the systems evolve downhill in free energy, in the direction of the gradient, within the relevant metric. Further, the probability of large deviations from the most likely hydrodynamic behaviour~\cite{bertini0102,bertini09,touchette-review} can be related to a geometrical ``action functional'' which also depends only on the free energy of the system and the relevant metric structure. The conclusion is that the theory of fluctuating hydrodynamics arises from a combination of a thermodynamic free energy functional, and a geometric structure that determines its dynamical evolution.  The relevant geometric structures are based on the Wasserstein distance~\cite{villani-book}, and have a physical interpretation in terms of a cost that is required to transport density through the system.

Several aspects of the situation that we present have been noted in existing work.  Formulae for large deviations in diffusive systems have been discussed in detail by Bertini~\emph{et al.}~\cite{bertini0102,bertini09,bertini14}, who interpreted the most likely path in the system as the solution to a Hamilton-Jacobi equation.  The path integral analyses that we will use to make the connection between geometrical structure and large deviations are standard, as summarised (for example) in~\cite{TKL}. In mathematics, connections between the Wasserstein geometry and hydrodynamic limit equations have been investigated~\cite{otto2001}, and generalisation of these results to large deviations (and fluctuating hydrodynamics) have also been discussed~\cite{adams2011,adams2013}.  Our main aim here is to interpret these mathematical results in a physical setting, which we accomplish by means of path integral methods.  The result is that the action functionals that appear in path integrals for diffusive systems can be given a geometrical interpretation in terms of steepest descent paths in the space of density profiles, providing an intuitive interpretation of these action functionals and (some of) their symmetry properties.

The form of the paper is as follows.  In Section~\ref{sec:models}, we discuss the diffusive systems to which our results apply, and we summarize their most relevant properties.  In Section~\ref{sec:geom}, we describe the Wasserstein geometry, and we show how it provides a geometrical interpretation of several results for the hydrodynamic limit of a system of non-interacting particles.  Then, in Section~\ref{sec:gen}, we discuss how the Wasserstein geometry can be modified such that it applies to broad class of diffusive systems.  We draw our conclusions and summarize outstanding issues in Section~\ref{sec:conc}.

\section{Model systems}
\label{sec:models}

\subsection{Free particle diffusion}

The results that we describe here are relevant for a range of model systems that include (for example) free-particle diffusion and the symmetric exclusion process.  In all cases, the behaviour
on hydrodynamic scales is captured by a locally-conserved density $\rho(\bm{r},t)$.  The models also exhibit microscopic reversibility, which
typically arises from a detailed balance relation at the level of the microscopic dynamics.  The simplest model system contains $N$ non-interacting
particles diffusing in $d$-dimensional space.  Starting from the master equation for an appropriate lattice model, standard methods leads to a path integral representation of the dynamics, valid on length scales much larger than the lattice spacing.  The construction of this path integral is outlined in~\ref{app:pathint}.  It is convenient to define $\varphi(\bm{r},t)=\frac1N \rho(\bm{r},t)$: since the system contains $N$ particles, we have $\int\!\mathrm{d}\bm{r} \varphi(\bm{r})=1$. (Here and in the following, $\bm{r}$-integrals run over the entire space of interest.)
The result (for large $N$) is that the expectation value of any density-dependent observable may be written as
\begin{equation}
\langle O \rangle = \frac{1}{Z} \int\! {\cal D}\varphi {\cal D}\hat{\rho}\, O[\rho] \exp\left[- \int\!\mathrm{d}t \int\! \mathrm{d}\bm{r} \, 
L(\varphi,\partial_t\varphi,\hat\rho,\nabla\hat\rho) \right]
\label{equ:O-pathint-free}
\end{equation}
with the Lagrangian
\begin{equation}
L(\varphi,\partial_t\varphi,\hat\rho,\nabla\hat\rho) = \hat\rho\partial_t\varphi + D\nabla\varphi\cdot\nabla\hat\rho - D\varphi|\nabla \hat\rho|^2 
\label{equ:lag-free}
\end{equation}
and $Z=\int\! {\cal D}\varphi {\cal D}\hat{\rho}\, \exp[-\int\!\mathrm{d}t \mathrm{d}\bm{r} \, L]$ for normalisation.
Here and throughout, we use $\langle\cdot\rangle$ to represent an average (or expectation value), 
and $\int{\cal D}\varphi{\cal D}\hat\rho$ to indicate a functional integral over paths $\varphi(\bm{r},t)$ and `response functions' $\hat\rho(\bm{r},t)$,
subject to the boundary conditions that are relevant for the physical situation of interest.   

The equilibrium state of this model has particle positions distributed independently and uniformly throughout the system: the microscopic dynamics obey detailed balance with respect to this distribution.  In terms of the density $\varphi$, the free energy that corresponds to this distribution is
\begin{equation}
\beta F[\rho] = \int\mathrm{d}\bm{r}\, \varphi [\log(\varphi\lambda^d) - 1]
\label{equ:F-free}
\end{equation}
where $\beta$ is an inverse temperature, and $\lambda$ a constant with units of length.  As expected, $F[\varphi]$ is the configurational part of the free energy of an ideal gas (which is purely entropic in this case).

At the level of the Lagrangian (\ref{equ:lag-free}), the time-reversal symmetry (detailed balance) of the model is not immediately apparent.  To reveal this symmetry, one should notice that
$\nabla\varphi = \varphi\nabla\frac{\delta}{\delta\varphi}(\beta F)$, and then make the change of variables $\varphi_{\rm TR}(\bm{r},t) = \varphi_{\rm TR}(\bm{r},-t)$, $\hat\rho_{\rm TR}(\bm{r},t) = -\hat\rho_{\rm TR}(\bm{r},-t) + \frac{\delta (\beta F)}{\delta\varphi}(\bm{r},-t)$~\cite{TKL}.  It follows that
\begin{eqnarray}
\fl\int_{-\tau}^{\tau}\mathrm{d}t\int\mathrm{d}\bm{r} L(\varphi_{\rm TR},\partial_t\varphi_{\rm TR},\hat\rho_{\rm TR},\nabla\hat\rho_{\rm TR})
&=& \beta F[\varphi(-\tau)] - \beta F[\varphi(\tau)] \nonumber\\ & & +\int_{-\tau}^\tau\mathrm{d}t\int\mathrm{d}\bm{r} L(\varphi,\partial_t\varphi,\hat\rho,\nabla\hat\rho)  .
\label{equ:time-rev}
\end{eqnarray}
This symmetry property of the path integral corresponds to the time-reversal symmetry of the underlying particle model.
One of the outcomes of the analysis presented here is that while this construction may seem both complicated and rather arbitrary,  it has a straightforward geometrical interpretation, within the Wasserstein metric: see Section~\ref{sec:rel} below.

\subsection{Other diffusive systems}

As well as this simple system of non-interacting particles, the results we will discuss also apply in a more general setting. 
We focus on hydrodynamic limits: Imagine observing a particle system on a large length scale, so that the fundamental
particles are no longer visible, and it is convenient to think in terms of a smooth density profile $\rho(\bm{r})$.  To take
the limit of a large observation scale, it is mathematically convenient to rescale co-ordinates and instead consider a fixed region 
of space in which the number of particles $N\to\infty$.  To enable this, we again define a rescaled density $\varphi(\bm{r},t) = \rho(\bm{r},t)/N$, so
that the hydrodynamic limit is $N\to\infty$ at fixed $\varphi$.  
Then the class of systems that we consider have path integral representations of the form
\begin{equation}
\langle O \rangle = \frac{1}{Z} \int\! {\cal D}\varphi {\cal D}\hat{\rho}\, O[\varphi] \exp\left[- N\int\!\mathrm{d}t \int\! \mathrm{d}\bm{r} \, 
L_m(\varphi,\partial_t\varphi,\hat\rho,\nabla\hat\rho) \right]
\label{equ:O-pathint}
\end{equation}
with the general Lagrangian
\begin{equation}
L_m(\varphi,\partial_t\varphi,\hat\rho,\nabla\hat\rho) = \hat\rho\partial_t \varphi + m(\varphi)D\nabla\hat\rho\cdot\nabla\frac{\delta}{\delta \varphi}\beta F[\varphi] - m(\varphi)D|\nabla \hat\rho|^2 .
\label{equ:lag-m}
\end{equation}
Here, $F[\varphi]$ is the free energy (per particle) of the diffusive system, and $m(\varphi)$ is a  density-dependent local mobility.  If we take  $F$ as in (\ref{equ:F-free}) and $m=\varphi$, we recover the free-particle model.  However, the class of models described by (\ref{equ:O-pathint}) also includes  non-interacting particles diffusing in a potential, the symmetric simple exclusion process (SSEP), and the Kipnis-Marchioro-Presutti (KMP) model: we discuss these specific cases in Section~\ref{sec:other-models} below.  Our results may be further generalised to cases where the mobility $m$ depends explicitly on $\bm{r}$, or to cases of anisotropic mobility (in which case $m$ becomes a $d\times d$ matrix).  It may easily be verified that systems described by (\ref{equ:O-pathint}) all have the time-reversal symmetry (\ref{equ:time-rev}): they have time-reversal symmetric (equilibrium) steady states with free energy $F$.

Note that we have taken $D$ to be a simple constant that sets the relative scaling space and time.  Alternatively we could incorporate the $\varphi$-dependence of $m$ into a $\varphi$-dependent diffusion constant $m(\varphi)D\to D(\varphi)$ as in~\cite{bertini14}: here, we anticipate that $m(\varphi)$ will have a geometrical interpretation (independent of time), so we separate it from the dynamical parameter $D$.

\subsection{Saddle point analysis and large deviation functional}

We now concentrate on the path integral (\ref{equ:O-pathint}), in the hydrodynamic (large-$N$) limit.    
In this limit, the integrand in (\ref{equ:O-pathint}) becomes sharply peaked about its maximal value.
Assuming that $O$ does not depend too strongly on $N$, one has
\begin{equation}
\langle O \rangle = O[\varphi^*],
\label{equ:O-rho-star}
\end{equation}
where $\varphi^*(\bm{r},t)$ is the path that maximises the exponential in (\ref{equ:O-pathint}), subject to any imposed boundary conditions.
We first maximise the exponential over the response field $\hat\rho$, for which the Euler-Lagrange
equation is
\begin{equation}
\partial_t \varphi = D \nabla\cdot m \left[ \nabla\frac{\delta}{\delta \varphi}\beta F - 2\nabla \hat\rho \right]  .
\label{equ:EL-rhohat}
\end{equation}
If we denote the ($\varphi$-dependent) solution of this equation by $\hat{\rho}^*$ and 
substitute into (\ref{equ:O-pathint}), 
the argument of the exponential reduces to $-N{\cal S}_m[\varphi]/2$ with
\begin{equation}
{\cal S}_m[\varphi] =  2 \int\mathrm{d}t\int\mathrm{d}\bm{r}\, m D \left| \nabla\hat\rho^* \right|^2.
\label{equ:S-rhostar}
\end{equation}
(The factor of $2$ is incorporated into ${\cal S}_m$ for later convenience.)
Minimising ${\cal S}_m$ then gives the most likely paths $\varphi^*$ (clearly, paths with $\nabla\rho^*=0$ are minimisers of ${\cal S}_m$, but the existence of such paths may depend on the boundary conditions in the path integral).
Hence, one may calculate expectation values of the form of (\ref{equ:O-rho-star}).

To arrive at a stronger result, we restrict
%
%
the integral in the numerator of (\ref{equ:O-pathint}) to trajectories $\varphi$ that are close to some reference path $\varrho(\bm{r},t)$.  Integrating over the $\hat\rho$ field, one finds
that the probability of observing such a path satisfies
\begin{equation}
\lim_{N\to\infty} \frac1N \log \mathrm{Prob}[\varphi(\bm{r},t) \approx \varrho(\bm{r},t)] = -\frac{1}{2} {\cal S}_m[\varrho] .
\label{equ:large-dev}
\end{equation}
This is a large deviation principle~\cite{touchette-review}, 
which determines the probability of observing a non-typical trajectory in this system, as $N\to\infty$.  Such large deviation principles play 
a central role in several theories of hydrodynamic behaviour~\cite{kov89,eyink1,bertini14}.

Equations~(\ref{equ:O-rho-star}-\ref{equ:large-dev}) allow analysis of hydrodynamic limits in a range of diffusive systems.
However, the physical interpretation of some of the quantities that appear in these equations is rather opaque.
The central point of this paper is that the large deviation function ${\cal S}_m$ has a geometrical interpretation, as do
the paths that minimise this function.  Hence, this geometrical structure controls determines both the most likely paths in the hydrodynamic
limit of these systems, as well as the probabilities of large deviations from these paths.  As we shall see, it also clarifies the connection between the functional
$\cal S$, the free energy $F$, and the time-reversal symmetry (detailed balance) of the microscopic dynamics in these systems.

\section{Wasserstein geometry}
\label{sec:geom}

The geometrical setting that we consider is called the Wasserstein geometry, which defines a metric structure in the space of density profiles $\rho(\bm{r})$.  
For a detailed discussion, we refer the reader 
to~\cite{villani-book} (see also~\cite{otto2001,adams2011,adams2013}).  The `standard' Wasserstein geometry
corresponds to the case $m=\varphi$ in the notation of (\ref{equ:lag-m}): others cases correspond to modified geometrical structures, 
as discussed in Section~\ref{sec:gen}.
%

\subsection{Wasserstein distance}
\label{sec:dist}

We begin by defining the Wasserstein distance between two density profiles $\rho_1(\bm{r})$ and $\rho_2(\bm{r})$, assuming that these profiles represent the same number $N$ of particles, $\int\! \mathrm{d}\bm{r} \rho_1(\bm{r}) = \int\! \mathrm{d}\bm{r} \rho_2(\bm{r}) = N$. We motivate the definition by a construction that comes from the theory of optimal transport~\cite{villani-book}. Starting with the profile $\rho_1(\bm{r})$, suppose that we redistribute the particle density over the whole space, with the fraction of the density from $\bm{r}$ that moves to $\bm{r}'$ being $q(\bm{r}'|\bm{r})$.  Clearly $\int\! \mathrm{d}\bm{r}' q(\bm{r}'|\bm{r})=1$ for all $\bm{r}$.  Also, for this process to generate the profile $\rho_2(\bm{r})$, one requires that $\int\! \mathrm{d}\bm{r} q(\bm{r}'|\bm{r}) \rho_1(\bm{r})=\rho_2(\bm{r}')$.  If transporting density through the system requires some kind of `cost' that is proportional to the square of the distance moved, then the total cost of this redistribution is
\begin{equation}
  C[q] = \int\mathrm{d}\bm{r}\mathrm{d}\bm{r}' |\bm{r}-\bm{r}'|^2   q(\bm{r}',\bm{r})
  \label{equ:cost}
\end{equation}
where $q(\bm{r}',\bm{r})=q(\bm{r}'|\bm{r}) \rho_1(\bm{r})$.  The (squared) Wasserstein distance between $\rho_1(\bm{r})$ and $\rho_2(\bm{r})$ is then defined by
\begin{equation}
  d[\rho_1,\rho_2]^2  = \inf_{q(\bm{r}',\bm{r})} C[q]
  \label{equ:d-cost}
\end{equation}
where the infimum (minimum) is taken over all distributions $q$ which satisfy the constraints $ \int\! \mathrm{d}\bm{r}' q(\bm{r}',\bm{r})=\rho_1(\bm{r})$ and $ \int\! \mathrm{d}\bm{r} q(\bm{r}',\bm{r})=\rho_2(\bm{r}')$.  That is, $q$ is a distribution, whose marginals are $\rho_1$ and $\rho_2$.  The Wasserstein distance is sometimes called an `earth-movers' distance, since it reflects the distance by which density must be transported, in which case $q(\bm{r}'|\bm{r})$ is called a ``transport plan''.

\subsection{Path lengths and path energies}
  
It is useful to consider continuous paths in the space of density profiles, represented as $\rho(\bm{r},s)$, where $s$ is a progress co-ordinate along the path.  
We suppose that $\rho(\bm{r},s)$ represents advection by an $s$-dependent velocity field $\bm{v}(\bm{r},s)$, so that 
$
\partial_s \rho = -\nabla\cdot( \rho\bm{v} )
$.
This advective process transports the density from an initial profile $\rho(\bm{r},a)$ to a final one $\rho(\bm{r},b)$.  Discretising the path into segments of length $\delta s$, 
we calculate the cost $\delta C$ of each segment by taking $q(\bm{r},\bm{r}')=\delta(\bm{r}-\bm{r}'-\bm{v}(\bm{r})\delta s)\rho(\bm{r})$ where $\rho(\bm{r})$ is the profile at the start of the segment.  The resulting cost is $\delta C = \int\mathrm{d}\bm{r} \rho |\bm{v}|^2 \delta s^2$, and we sum over the segments to obtain 
$
  {\cal E}_v[\rho(\bm{r},s),\bm{v}(\bm{r},s)] = \lim_{\delta s\to0} \frac{1}{\delta s}\sum_k \delta C_k =  \int_a^b\mathrm{d}s    \int\mathrm{d}\bm{r}\, \rho |\bm{v}|^2$,
which depends both on the path $\rho(\bm{r},s)$ and the velocity field $\bm{v}(\bm{r},s)$.  
For any path, this cost is minimised by a potential flow $\bm{v}=-\nabla\Phi$~\cite{benamou}.  We therefore define a new cost
\begin{equation}
{\cal E}[\rho(\bm{r},s)] = \int_a^b\mathrm{d}s  \int\mathrm{d}\bm{r}\, \rho |\nabla \Phi|^2
\label{equ:path-cost}
\end{equation}
where 
$\nabla\Phi$ satisfies
\begin{equation}
\partial_s \rho = \nabla\cdot( \rho \nabla \Phi).
\label{equ:rho_s_Phi}
\end{equation}
In contrast to ${\cal E}_v$, the cost ${\cal E}$ depends only on the path $\rho(\bm{r},s)$.

It then follows~\cite{benamou} that the Wasserstein distance between two profiles $\rho_1(\bm{r})$ and $\rho_2(\bm{r})$ can be obtained by finding the path of minimal cost $\cal E$ between these points.  The result is that 
\begin{equation}
  d[\rho_1,\rho_2]^2=(b-a) \inf_{\rho(\bm{r},s)} 
{\cal E}[\rho]
  \label{equ:d-path-cost}
\end{equation} 
where the minimisation is subject to $\rho(\bm{r},a)=\rho_1$ and $\rho(\bm{r},b)=\rho_2$.  Note that the cost $\cal E$ was defined by summing the costs $\delta C$ of each segment along the path: from (\ref{equ:d-cost}), the cost $\delta C$ is a squared distance so $\cal E$ corresponds to a sum of squared segment lengths.  However, the distance $d$ should be equal to the sum of the segment lengths themselves: the content of (\ref{equ:d-path-cost}) is that the infimum (minimum) is obtained when all segments are of equal length, in  which case the distance $d$ can be inferred from the (rescaled) sum of the  squared segment lengths.  The factor of $(b-a)$ in (\ref{equ:d-path-cost}) and the factor of $\delta s$ in the definition of ${\cal E}_v$ implement the required rescaling.  Mathematically, this relation follows from a Cauchy-Schwartz inequality between path actions and path lengths: see~\ref{app:metric}.

Hence,
the path $\rho(\bm{r},s)$ that realises the minimum in (\ref{equ:d-path-cost}) is the geodesic (shortest path) connecting $\rho_1$ and $\rho_2$, in the Wasserstein geometry.  The metric structure that underlies these results is discussed in more detail in \ref{app:metric}.  We have presented a heuristic argument for (\ref{equ:d-path-cost}), but this result has been shown rigorously~\cite[Proposition 1.1]{benamou}.  

\subsection{Steepest descent with respect to a free energy}
\label{subsec:descent}

A central conclusion of this study is that diffusive processes are governed by two factors: the geometrical structure of the space of density profiles, and the free energies of these profiles.  We therefore associate to each profile $\rho(\bm{r})$ a free energy $F[\rho]$.  In this section, we focus on the case where $F$ is given by (\ref{equ:F-free}), which corresponds to a microscopic model of non-interacting particles.

Given a free energy and a distance measure (metric), it is natural to define a steepest descent process.  Paths of steepest descent may be constructed by a discrete-time process \cite{jko}: on each time step, a system with profile $\rho_t(\bm{r})$ evolves to the profile $\rho_{t+\delta t}(\bm{r})$ which minimises
\begin{equation}
\label{jko}
S[\rho_{t+\delta t}] := \frac{1}{2D\delta t} d[\rho_t,\rho_{t+\delta t}]^2 + \beta F[\rho_{t+\delta t}] .
\end{equation}
Here $\delta t$ is the (small) time increment associated with each time step and $D$ a constant that sets the units of time.  (We note that $d$ has units of length and $D t$ has units of $\mathrm{length}^2$, while $S$ and $\beta F$ are dimensionless.)  To make contact with the previous subsection, we identify $D t$ with the progress variable $s$.
Assuming that $\delta t$ is small, we then use (\ref{equ:path-cost}) and  (\ref{equ:d-path-cost}) to approximate $S[\rho_{t+\delta t}]$ by
\begin{equation}
  \label{eq:jkowass}
   \frac {D \delta t} {2} \int\rho_t |\nabla \Phi|^2 \, \mathrm{d}\bm{r} 
                                  + \beta F[\rho_{t+\delta t}] 
\end{equation}
where the dimensionless field $\Phi(\bm{r},t)$ solves 
\begin{equation}
\partial_t \rho = \nabla\cdot\left( \rho D \nabla\Phi\right).
\label{equ:dtrho}
\end{equation}

We now remark that minimising $S$ is equivalent to minimising $\tilde S = S - \min S$; the latter is a nonnegative functional. We therefore minimise (\ref{eq:jkowass}) over $\rho_{t+\delta t}$, using a Lagrange multiplier $\lambda=\lambda(\bm{r})$ to enforce the constraint (\ref{equ:dtrho}).  The quantity to be extremised (over $\lambda,\Phi,\rho_{t+\delta t}$) is 
\begin{equation} \fl 
\frac {D \delta t} {2} \int \rho_t |\nabla \Phi|^2 \mathrm{d}\bm{r} + \beta F[\rho_{t+\delta t}] - \int \lambda[ \rho_{t+\delta t} - \rho_t - D \delta t \nabla \cdot (\rho \nabla \Phi)] \mathrm{d}\bm{r} .
\end{equation} 
Extremising over $\rho_{t+\delta t}$ yields $\lambda = \beta\delta F/\delta \rho$ and extremising over $\Phi$ gives $(D\delta t)\nabla\cdot(\rho\nabla\Phi) = (D\delta t)\nabla\cdot(\rho\nabla\lambda) $.  We solve these equations by taking $\Phi=\lambda=\beta\delta F/\delta \rho$: this means that (\ref{equ:dtrho}) becomes the equation of steepest descent for $\rho$ [see (\ref{equ:rho-gradflow}) below].  It remains to compute the minimal value of $S$: we write $\beta F[\rho_{t+\delta t}] \approx \beta F[\rho_t] + (\rho_{t+\delta t} - \rho_t)\beta\delta F/\delta \rho$ and the constraint (\ref{equ:dtrho}) implies that $\rho_{t+\delta t} - \rho_t = D \delta t \nabla \cdot (\rho \nabla \Phi)$.  Substituting for $\beta F[\rho_{t+\delta t}]$ and $\Phi$ in (\ref{jko}) gives $\min_{\rho_{t+\delta t}} S = \beta F[ \rho_t ] 
             - \frac{D \delta t}2 \int_{\bm r} \rho_t  | \nabla (\delta \beta  F[\rho_t] / \delta \rho) |^2$.  Hence
\begin{equation}
  \label{eq:jko-expand}
  \fl
  \tilde{S}[\rho_{t+\delta t}] = 
  \frac {D \delta t} {2} \int \rho_t |\nabla \Phi|^2  \mathrm{d}\bm{r}
             + \beta F[ \rho_{t+ \delta t} ]  - \beta F[ \rho_t ]
             + \frac {D \delta t} 2 \int \rho_t \left|\nabla\frac{\delta }{\delta \rho}\beta F[\rho_t] \right|^2 \mathrm{d}\bm{r}.
\end{equation}

Recall that $\tilde S[\rho_{t+\delta t}]$ measures the difference between $S[\rho_{t+\delta t}]$ and its minimal value: it is equal to $S[\rho_{t+\delta t}] - S[\rho^*]$, where $\rho^*$ is the profile that would be chosen in a steepest descent process.  Moving from this discrete-time construction to a cost functional for continuous
 paths, and using the path cost $ {\cal E}[\rho] = \int\mathrm{d}t \int\mathrm{d}\bm{r} \rho |\nabla \Phi|^2$ [where $\Phi$ solves (\ref{equ:dtrho})], one arrives at
 \begin{equation}
\fl
{\cal S}[\rho(\bm{r},s)] =  \frac{D}{2} {\cal E}[\rho] +  \beta F[\rho(\bm{r},b)] - \beta F[\rho(\bm{r},a)] + \frac{D}2 \int_a^b\mathrm{d}t  \int\mathrm{d}\bm{r}\, \rho  \left|\nabla\frac{\delta }{\delta \rho}(\beta F)\right|^2 .
\label{equ:cost-gradflow2}
\end{equation}
Finally, noting that $  F[\rho(\bm{r},b)] - F[\rho(\bm{r},a)]= \int_a^b\mathrm{d}t (\delta F/\delta \rho) \partial_t \rho$, one arrives at the path functional  for steepest descent in the Wasserstein metric:
\begin{equation}
{\cal S}_{\rm W}[\rho(\bm{r},t)] =  \frac12 \int_a^b\mathrm{d}t  \int\mathrm{d}\bm{r}\, \rho D \left|\nabla \left(\Phi - \beta\frac{\delta F}{\delta \rho}\right)\right|^2
\label{equ:cost-gradflow}
\end{equation}

The meaning of (\ref{equ:cost-gradflow}) is that if $-D\nabla\Phi$ is the velocity field that advects the underlying particle density, the effect of the free energy $F$ is to bias this advection.  Also, the paths that minimise $\cal S$ (which are paths with ${\cal S}=0$) satisfy
\begin{equation}
\partial_t \rho = \nabla\cdot\left[\rho D\nabla\frac{\delta }{\delta \rho}(\beta F)\right]  =: - D\, \mathrm{grad}^{\rm W}(\beta F) . 
\label{equ:rho-gradflow}
\end{equation}
Here, the `gradient operator' $\mathrm{grad}^{\rm W}$ for the Wasserstein metric is defined by its action on a functional $\Psi[\rho(\bm{r})]$, as $\mathrm{grad}^{\rm W}(\Psi) = -  \nabla\cdot\left(\rho\nabla\frac{\partial}{\partial  \rho}\Psi\right)$. So (\ref{equ:rho-gradflow}) describes steepest descent with respect to the free energy $F$ in the Wasserstein geometry. This formulation is analogous to steepest descent in Euclidean space, where $\gamma\partial_t \bm{x} = - \nabla_{\bm x} E$, in which $\gamma$ is a friction constant and $E$ a potential energy function. While this analogy looks at the moment formal and possibly arbitrary, since it is based on the definition of $\mathrm{grad}^{\rm W}$, it can be shown that the Wasserstein geometry indeed defines an infinite-dimensional geometric structure where $\mathrm{grad}^{\rm W}$ plays the part of a gradient. Elements of this theory are sketched in~\ref{app:metric}.  

\subsection{Relation to diffusive systems: saddle point trajectories and large deviations}
\label{sec:rel} 

We are now in a position to connect this geometrical construction to the results of Section~\ref{sec:models}.  The results obtained so far correspond to the case $m=\varphi$ in that Section.
The first thing to note is that the most likely paths $\varphi^*$ in (\ref{equ:O-rho-star}) are equal (up to a factor of $N$) to the solutions of the steepest descent equation (\ref{equ:rho-gradflow}).  This follows because the most likely paths are those with $\nabla\hat\rho^*=0$ [see (\ref{equ:S-rhostar})], 
in which case setting $m=\varphi$ in (\ref{equ:O-rho-star}) yields (\ref{equ:rho-gradflow}).  
It is also instructive to substitute for $F$ using (\ref{equ:F-free}), in which case (\ref{equ:rho-gradflow}) reduces to the diffusion equation
\begin{equation}
  \label{hamster}
  \partial_t \rho = - D\, \mathrm{grad}^{\rm W}(\beta F) = \nabla\cdot\left[\rho D\nabla\frac{\delta }{\delta \rho}(\beta F)\right] = D\nabla^2\rho   .
\end{equation}
Thus, the trajectories that dominate the hydrodynamic limit of this system correspond to steepest descent paths of the free energy, within the Wasserstein metric~\cite{jko}.

In addition, the large deviation function ${\cal S}_m$ in (\ref{equ:S-rhostar}) corresponds (for $m=\varphi$) to the Wasserstein path action ${\cal S}_{\rm W}$
in (\ref{equ:cost-gradflow}).  To see this, note that since $\hat\rho^*$ in (\ref{equ:S-rhostar}) solves Eq.~(\ref{equ:EL-rhohat}) then (for $m=\varphi$) we can identify $\hat\rho = \frac12[\delta(\beta F)/\delta \varphi - \Phi]$ where $\Phi$ solves (\ref{equ:dtrho}).  Thus the action ${\cal S}_m$ that determines path probabilities in the large deviation principle (\ref{equ:large-dev}) is the same as the Wasserstein path action ${\cal S}_{\rm W}$.  That is, the Wasserstein metric and the free energy together specify not just the dominant hydrodynamic path but also the fluctuations about this path.

Finally, we note from (\ref{equ:cost-gradflow2}) that the cost of a path depends on the direction
with which the path is traversed: the free energy difference gives a contribution that is odd under time reversal while the first and last terms on the right-hand side of (\ref{equ:cost-gradflow2}) are both even under time-reversal.  The fact that the odd part of the cost is simply the free energy difference between start and end points is equivalent to the detailed balance symmetry of the microscopic model: for large $N$, if $\varphi_{\rm TR}$ is the time-reversed counterpart of a trajectory $\varphi$ that runs from $t=a$ to $t=b$ then one has from (\ref{equ:large-dev}) and (\ref{equ:cost-gradflow2}) that 
$\ee^{-\beta F(a)}\mathrm{Prob}[\varphi]=\ee^{-\beta F(b)}\mathrm{Prob}[\varphi_{\rm TR}]$.  The appearance of the free energy in (\ref{equ:cost-gradflow2}) may appear coincidental from the derivation given here but this is a general property of steepest-descent processes (gradient flows), which follows from the the definition of the path cost (\ref{equ:cost-gradflow}): see also~\ref{app:metric}.  Hence, if a system obeys a large deviation function of the form of (\ref{equ:large-dev}), where the action ${\cal S}$ corresponds to the path action for a steepest descent process, then one arrives at a detailed balance-like relation which relates the probability of trajectories to their time-reversed counterparts.  The generalisation of this result to systems without detailed balance would presumably result in  a fluctuation theorem similar to that of Crooks~\cite{crooks}: this would be an interesting direction for future study.

\subsection{Fluctuating hydrodynamics}

Since the Wasserstein metric provides a connection both to the most likely hydrodynamic path and to fluctuations about this path, it can also be used to interpret the theory if `fluctuating hydrodynamics' for diffusive systems~\cite{spohn83,TKL}.
Within this framework, one describes systems on macroscopic scales by Langevin equations, or stochastic partial differential equations.  For free particle diffusion, the relevant equation is
\begin{equation}
\partial_t \rho = D[ \nabla^2\rho + \nabla\cdot(\bm{\eta}\sqrt{\rho}) ]
\label{equ:FH}
\end{equation}
where $\bm{\eta}$ is a space-time white noise.  [That is, a Gaussian-distributed random function with $\langle \bm{\eta}\rangle=0$ and $\langle \eta^\mu(\bm{r},t) \eta^\nu(\bm{r}',t') \rangle = \delta^{\mu\nu} \delta(t-t') \delta(\bm{r}-\bm{r}')$, where $\mu,\nu$ label Cartesian components of the vector $\bm{\eta}$.]  In the hydrodynamic limit, it may be more convenient to write (\ref{equ:FH}) as $\partial_t \varphi = D\left[ \nabla^2\varphi +   \frac{1}{\sqrt{N}}\nabla\cdot(\bm{\eta}\sqrt{\varphi}) \right] $ to emphasise that effect of the noise becomes increasingly weak as the large-$N$ limit is approached.  Following the procedure of Martin-Siggia-Rose-DeDominicis-Janssen~\cite{msr,dedom,janssen}, one may show that equation (\ref{equ:FH}) is equivalent to the path-integral description (\ref{equ:O-pathint}): see Section 3 of Ref.~\cite{TKL} for a detailed discussion.

If the density profile at time $t$ is $\rho_1(\bm{r})=\rho(\bm{r},t)$, the Langevin equation (\ref{equ:FH}) specifies a probability distribution for the density a short time $\delta t$ later, $\rho_2(\bm{r})=\rho(\bm{r},t+\delta t$).  In the steepest descent case (where the noise $\bm{\eta}$ is absent), we recall from (\ref{jko}) that $\rho_2$ may be obtained by minimising $\frac12 d[\rho_1,\rho_2]^2+D \delta t \beta F[\rho_2]$.  To incorporate the effects of noise, we start from (\ref{equ:large-dev}) and time-discretise the action $\cal S$, from which we find that the probability distribution of $\rho_2$ is
\begin{equation}
p[\rho_2|\rho_1,\delta t] \propto \exp\left(  -\frac{1}{4 D\delta t} d[\rho_1,\rho_2]^2-\frac12\beta F[\rho_2]  \right).
\label{equ:prho2}
\end{equation}
[To be precise, $p[\rho_2]$ is a probability density: probabilities are obtained by functional integrals of the form $\int{\cal D}\rho_2\, p[\rho_2]$ 
where the functional integral runs over functions $\rho_2(\bm{r})$ with $\int\mathrm{d}\bm{r} \rho_2 = N$, so that the total density is conserved.  
To reiterate, the Langevin equation (\ref{equ:FH}) is equivalent to the path integral (\ref{equ:O-pathint}), and for large-$N$
the probabilities of trajectories in this system satisfy both (\ref{equ:large-dev}) and (\ref{equ:prho2}).  Thus, (\ref{equ:prho2}) is a large-$N$ result
for the stochastic evolution defined by (\ref{equ:FH}).  It means that the probability that a density profile $\rho_1$ evolves
into $\rho_2$ over a short time period $\delta t$ has a Gaussian dependence on the distance $d[\rho_1,\rho_2]$ and a simple
exponential dependence on the free energy of the final state $\beta F[\rho_2]$.  

It is useful to compare this result with the overdamped Langevin of a single particle in a potential $V(\bm{r})$.  If the position of the particle is $\bm{x}(t)$, one writes
\begin{equation}
\partial_t \bm{x} = D \beta\nabla V + \bm{\eta}_0
\end{equation}
where $\gamma$ is a friction constant, the noise $\bm{\eta}_0$ has covariance $\langle \eta^\mu_0(t) \eta_0^\nu(t') \rangle = 2D\delta^{\mu\nu} \delta(t-t')$, and $\beta$ is the inverse temperature.  This equation implies that if $\bm{x}(t)=\bm{x}_1$, then after a small time interval $\delta t$, the position of the particle $\bm{x}_2=\bm{x}(t+\delta t)$ is distributed as
\begin{equation}
p(\bm{x}_2|\bm{x}_1,\delta t) \propto \exp\left[-\frac{1}{4D\delta t}|\bm{x}_2-\bm{x}_1|^2 
- \frac12 \beta V(\bm{x}_2)\right].
\label{equ:px2}
\end{equation}
We again recover Gaussian dependence on the distance $|\bm{x}_2-\bm{x}_1|$ and exponential dependence on the free energy difference (here the free energy is given simply by the energy). Based on the similarity between (\ref{equ:prho2}) and (\ref{equ:px2}), we argue that (\ref{equ:FH}) is the natural formulation of overdamped dynamics in the Wasserstein geometry -- this is a geometrical interpretation of the theory of fluctuating hydrodynamics for this system. The  particular noise in (\ref{equ:FH}) has been connected to the Wasserstein geometry before on mathematical grounds \cite{max}; here we arrive at the same result with a very different argument, starting from particles and a path integral formulation.

\section{Generalisation to other systems}
\label{sec:gen} 

\subsection{Modified Wasserstein metric}

We have illustrated a connection between the Wasserstein distance as defined by (\ref{equ:d-cost}) and the hydrodynamic limit for diffusion of free particles.  [In the notation of Section~\ref{sec:models}, the results of Section~\ref{sec:geom} apply only in the case where $m=\varphi$ and $F$ is given by (\ref{equ:F-free}).]  To generalise this connection to the broader class of systems anticipated in (\ref{equ:O-pathint}), we define a generalised Wasserstein distance.
To this end, define a modified path cost by analogy with (\ref{equ:path-cost}): ${\cal   E}_\mob[\rho(\bm{r},s)] = \int_a^b\mathrm{d}s \int\mathrm{d}\bm{r}\, \mob |\nabla \Phi_\mob|^2$ where $\Phi$ solves $\partial_s \rho = \nabla\cdot(\mob\nabla\Phi_\mob)$. Then a construction of steepest descent trajectories as in Section~\ref{subsec:descent} yields a generalised action functional
\begin{equation}
  {\cal S}_\mob[\rho(\bm{r},t)] = \frac12 \int_a^b\mathrm{d}t\,    \int\mathrm{d}\bm{r}\, \mob D\left|\nabla\Phi_\mob -   \nabla\frac{\delta (\beta F)}{\delta \rho} \right|^2  ,
\label{equ:S-cost-sig}
\end{equation}
where  $\Phi_\mob$ solves
$\partial_t \rho = \nabla\cdot(D\mob\nabla\Phi_\mob)$.  Repeating the analysis of Sec.~\ref{sec:rel}, it is easily checked that this path action is equal to the large deviation function in (\ref{equ:large-dev}): it follows that the dominant hydrodynamic trajectories are therefore steepest descent processes of the free energy within the relevant metric, that a detailed balance symmetry holds at the macroscopic level, and that one may apply the theory of fluctuating hydrodynamics in this more general case too.  To make contact with previous work~\cite{bertini14}, it is useful to note that the operator $K(\rho)$ in~\cite{bertini14} is the metric tensor for this generalised Wasserstein geometry: see \ref{app:metric}.

The key results are therefore that the action ${\cal S}_m$ appearing in the large deviation principle (\ref{equ:large-dev}) is the relevant modified Wasserstein path action (\ref{equ:S-cost-sig}), and that the corresponding fluctuating hydrodynamic equation is
\begin{equation}
\partial_t \varphi = D\nabla \cdot  \left[ \mob\nabla\frac{\delta}{\delta\varphi}(\beta F) + \frac{1}{\sqrt{N}} (\bm{\eta}\sqrt{\mob}) \right].
\label{equ:FH-gen}
\end{equation}
The generalisation of (\ref{equ:prho2}) that is equivalent to this process is obtained by replacing the distance $d$ in (\ref{equ:prho2}) by the modified distance $d_{\mob}$.  We therefore interpret (\ref{equ:FH-gen}) as the natural generalisation of Langevin dynamics to the modified Wasserstein geometry.  The dominant hydrodynamic path for this system is obtained by dropping the noise term from (\ref{equ:FH-gen}), leading to $\partial_t\varphi=D\,\mathrm{grad}^{\rm W}_\mob (\beta F)$ where $\mathrm{grad}^{\rm W}_\mob (\cdot) = \nabla\cdot[\mob D \nabla\frac{\delta}{\delta   \varphi}(\cdot)]$ is the definition of the gradient operator within the modified Wasserstein geometry.

\subsection{Diffusive systems described by modified Wasserstein geometries}
\label{sec:other-models}

To illustrate the range of physical systems to which this analysis applies, we now describe three physical systems that are described by path integrals of the form of (\ref{equ:O-pathint}), and we discuss the metric structures associated with the hydrodynamic limits of this models. The large deviation functions and the fluctuating hydrodynamic equations for all these models are related to modified Wasserstein distances, as we have outlined here.  (We also note in passing that the case of additive noise, $\mob=1$, leads to model-B dynamics~\cite{hohenberg-halperin}.)

\subsubsection{Diffusion of free particles in a potential}

The simplest generalisation of free particle diffusion is to introduce a potential $V(\bm{r})$ that is smooth on the hydrodynamic scale.  In this case the free energy is simply
\begin{equation}
\beta F[\rho] = \beta F_{\rm id}[\rho] + \int\mathrm{d}\bm{r}\, \rho\beta V ,
\label{equ:FV}
\end{equation} where $F_{\rm id}[\rho]$
is the non-interacting free energy given in (\ref{equ:F-free}).   
Constructing the path integral as in the free-particle case, one obtains a Lagrangian of the form (\ref{equ:lag-m}): 
\begin{equation}
L  = \hat\rho\partial_t \varphi + \nabla\hat\rho\cdot D(\nabla\varphi+\varphi\beta\nabla V) - D\varphi|\nabla \hat\rho|^2 .
\end{equation}
Comparing with (\ref{equ:lag-m}), one identifies $\mob=\varphi$ as in the free particle case, and $\mob\nabla\frac{\delta}{\delta\varphi}(\beta F)=\nabla\varphi+\varphi\beta\nabla V$, as required.  Hence, modifying free-particle diffusion by including an external potential preserves the connection to the Wasserstein geometry that was already identified in Sec.~\ref{sec:rel}.

\subsubsection{Symmetric exclusion process}

In the simple symmetric exclusion process (SSEP), particles hop between the sites of a lattice, subject to the constraint that at most one particle may occupy any site.  This model is simple to define but interactions between particles are strong, and the behaviour of the system is richer as a result.  The free energy for the SEP (on the hydrodynamic scale) is
\begin{equation}
\beta F = \int\mathrm{d}\bm{r}\, [\varphi\log\varphi + (1-\varphi)\log(1-\varphi)] ,
\label{equ:F-sep}
\end{equation}
where $\varphi$ is the local density, rescaled to lie between zero and unity, with $\varphi(\bm{r})=1$ corresponding to almost all sites being occupied in the vicinity of the point $\bm{r}$.

In the hydrodynamic limit, the system can be described~\cite{TKL} by a path integral of the form of (\ref{equ:O-pathint}), with
\begin{equation}
L  = \hat\rho\partial_t \varphi + \nabla\hat\rho\cdot (D\nabla\varphi) - D\varphi(1-\varphi)|\nabla \hat\rho|^2 .
\end{equation}
Comparing with (\ref{equ:lag-m}), we identify $\mob=\varphi(1-\varphi)$, so consistency between (\ref{equ:lag-m}) and (\ref{equ:F-sep}) requires $\varphi(1-\varphi) \nabla\frac{\delta}{\delta\varphi}(\beta F)=\nabla\varphi$.  This may be verified from (\ref{equ:F-sep}). Hence, a large deviation principle of the form of (\ref{equ:large-dev}) applies to the SSEP as well as to non-interacting systems, where the large deviation function is the action functional for the relevatn modified Wasserstein metric. (Properties of this large deviation function have also been discussed extensively by Bertini~\emph{et al.}~\cite{bertini0102,bertini09}).  For a fixed initial condition, the most likely trajectories in the SSEP are given by $\partial_t \varphi = D\nabla\cdot(\mob\nabla\frac{\delta}{\delta\varphi}\beta F[\varphi])=D\nabla^2\varphi$, corresponding to steepest descent in the relevant metric. We emphasise that while we recover the same diffusion equation as in the free-particle case, both the free energy $F$ and the function $\mob$ that determines the geometry are different, so the fluctuations about the most likely path differ strongly between the SEP and the free particle model. Specifically, the equation of fluctuating hydrodynamics for the symmetric exclusion process is
\begin{equation}
  \partial_t \varphi = D[ \nabla^2\varphi + \nabla\cdot(\bm{\eta}\sqrt{\varphi(1 - \varphi)/N}) ] ,
\end{equation}
which differs from (\ref{equ:FH}). Here, $\bm{\eta}$ is a space-time white noise, as above.

\subsubsection{Kipnis-Marchioro-Presutti (KMP) model}

The KMP model~\cite{kmp} was developed as a model for heat conduction and has been studied extensively as a model diffusive system~\cite{bertini-kmp05,TKL,lecomte10}.  In this model, $\varphi$ denotes a local energy density.  The model is defined on a $d$-dimensional lattice, and at each time step, the energies of two neighbouring sites are redistributed between those sites.  In the equilibrium state, the energy of each site is exponentially distributed, leading to a free energy on the hydrodynamic scale given by~\cite{bertini-kmp05}
\begin{equation}
\beta F = \int\mathrm{d}\bm{r}\, [  \beta\varphi - 1 - \log(\beta\varphi)].
\label{equ:F-kmp}
\end{equation}
Constructing the path integral for the dynamical evolution of the model, the Lagrangian is~\cite{TKL,lecomte10}
\begin{equation}
L  = \hat\rho\partial_t \varphi + \nabla\hat\rho\cdot (D\nabla\varphi) - D\varphi^2|\nabla \hat\rho|^2 ,
\end{equation}
so we identify $\mob=\varphi^2$. It is easily verified from (\ref{equ:F-kmp}) that $\varphi^2\nabla\frac{\delta}{\delta   \varphi}(\beta F) = \nabla\varphi$, as required for consistency with (\ref{equ:lag-m}).  Thus the deterministic (hydrodynamic) equation is again $\partial_t\varphi=D\nabla^2\varphi$, as for free particles and for the SEP, but in the KMP model this limit equation arises from a different combination of a free energy and a (generalised Wasserstein) geometrical structure.

\section{Conclusion}
\label{sec:conc}

We have shown how large deviations in the hydrodynamic theory of several diffusive systems can be interpreted geometrically, in terms of the Wasserstein geometry and its generalisations.  In particular, the most likely trajectories for these systems in the hydrodynamic limit are given by steepest descent of the free energy, in the appropriate metric.  We also argued that the equations of fluctuating hydrodynamics are the natural generalisations of Langevin dynamics, within this geometrical structure.  The relation between the large deviations of the time-dependent density and steepest descent processes clarifies the decomposition of the large deviation function into parts that are even and odd under time reversal, which can be traced back to detailed balance properties of the original stochastic processes.  It would be interesting to understand whether other properties of these systems such as responses to boundary driving~\cite{bertini09,TKL} or large deviations of time-integrated quantities~\cite{imparato09,lecomte10} can be related to properties of the Wasserstein geometry (see~\cite{mouhot} for nonlinear diffusion with inhomogeneous boundary data).
 
More generally, for models where the general structure of Section~\ref{sec:gen} applies, all hydrodynamic properties are determined by the (thermodynamic) free energy functional $F$, and the geometrical function $\mob$, which determines the `cost' of density redistribution in the system.  The idea that the dynamical evolution of stochastic models arises from a combination of a free energy and a metric structure has been discussed in other contexts too~\cite{whitelam-metric04}: the freedom to choose different metrics while preserving the same free energy functional means that systems with the same thermodynamic properties can have very different dynamical behaviour~\cite{ritort-sollich,jacquin12-arxiv}.  

It would be interesting to find other examples of stochastic processes whose behaviour can be analysed using metric structures in infinite-dimensional spaces, such as the space of density profiles considered here. One particular aspect here is the derivation of systems driven by energy and entropy. In particular, there is a rich class of equations of fluctuating hydrodynamics with the dissipative part being of the form (\ref{equ:FH-gen})~\cite{eyink1,eyink2}.

\ack We would like to thank Alan McKane, Vivien Lecomte, Mark A. Peletier, Max von Renesse, Tim Rogers, Peter Sollich, and Fred van Wijland for helpful discussions and comments on the manuscript.  RLJ was supported by the EPSRC through grant EP/I003797/1.  JZ gratefully acknowledges financial
support from the EPSRC (EP/K027743/1) and the Leverhulme Trust (RPG-2013-261).

\begin{appendix}

\section{Path integral construction}
\label{app:pathint}



In this section, we give a very brief review of the construction of the path integral expression (\ref{equ:O-pathint}), in order
that the presentation of this paper be as self-contained as possible, and to emphasise that this path integral comes from a microscopic
description of a specific particle system.
The microscopic model involves particles hopping on a (hyper)cubic 
lattice in $d$-dimensions.  The lattice spacing is $l_0$ and particles hop independently, with a rate $D/l_0^2$ for hopping
along each available bond on the lattice.  The number of particles on site $i$ is $n_i$, and a configuration $\CC$ is specified
by the values of all the $n_i$.  

Let $P(\CC,t)$ be the probability of finding the system in configuration $\CC$ at time $t$.  This quantity evolves in time by
a master equation.  To obtain a representation of this equation, it is useful~\cite{doi-peliti} 
to write the probability distribution as a high-dimensional
vector $|P\rangle = \sum_\CC P(\CC,t) |\CC\rangle$.  Within this vector space, the operator $a^\dag_i$ acts on $|\CC\rangle$ by adding
 a particle
to site $i$, while $a_i$ multiplies by $n_i$ before removing a particle from site $i$ (the occupancies of all sites $j\neq i$ are unchanged).
Hence $a_i^\dag a_i|\CC\rangle = n_i|\CC\rangle$ and one has also the commutation
relation $[a_i,a_j^\dag]\equiv a_i a_j^\dag - a_j^\dag a_i = \delta_{ij}$. It is also useful to denote the state with no 
particles at all by $|0\rangle$.
Then the master equation may be written as~\cite{doi-peliti}
\begin{equation}
\partial_t |P\rangle = \mathbb{W} |P\rangle
\end{equation}
with 
\begin{equation}
\mathbb{W} = D l_0^{-2} \sum_{\langle ij\rangle}  (a_i^\dag - a_j^\dag)(a_j - a_i) 
\end{equation}
where the sum runs over pairs of nearest neighbours on the lattice.

The formal solution of the master equation is $|P(t)\rangle = {\rm e}^{\mathbb{W}t}|P(0)\rangle$.
If we assume for convenience that the operator $O$ in (\ref{equ:O-pathint}) depends only on the density at a single
time $t_1$ then it may be shown that $\langle O \rangle = \langle 0 | {\rm e}^{\sum_i a_i} \hat{O} {\rm e}^{\mathbb{W}t_1}|P(0)\rangle$
where $\hat{O}$ is an operator that depends on the density $\hat{\rho} = \sum_i a_i^\dag a_i$ in the same way that $O$ depends on $\rho(t_1)$.
Generalisation to other observables $O$ is straightforward but we omit it, for brevity.

To make further progress, one makes a time discretisation, writing ${\rm e}^{\mathbb{W}t_1} = \prod_{k=1}^M {\rm e}^{\mathbb{W}\delta t}$ 
with $\delta t = t_1/M$.  One also requires a formula for the identity operator~\cite{doi-peliti}:
\begin{equation}
\hat{1} = \int\left[\prod_i\mathrm{d}(z_i,z_i^*)\right]\, \ee^{-\sum_i z_i z_i^*} \ee^{\sum_i z_i a_i^\dag}|0\rangle\langle 0| \ee^{\sum_i z_i^* a_i}
\end{equation}
where $z^*_i$ is the complex conjugate of $z_i$, and the notation $[\prod_i\mathrm{d}(z_i,z_i^*)]$ means that each $z_i$ is integrated over the entire complex plane, with one integration variable $z_i$ for each
site on the lattice.  Inserting this resolution of the identity between each factor of ${\rm e}^{\mathbb{W}\delta t}$ in the representation
of ${\rm e}^{\mathbb{W}t_1}$, one arrives at
\begin{equation}
\fl
\langle O \rangle = \int \left[ \prod_{ik}\mathrm{d}(z_{ik},z_{ik}^*) \right] 
\langle 0 | {\rm e}^{\sum_i a_i} \hat{O} \ee^{\sum_i z_{iM} a_i^\dag}|0\rangle \ee^{-S_0} \ee^{-\sum_i z^*_{i0} z_{i0}}
\langle 0| \ee^{\sum_i (z_{i0})^* a_i}|P(0)\rangle
\label{equ:pi1}
\end{equation}
where
\begin{equation}
\ee^{-S_0} = \prod_{k=1}^M  \ee^{-\sum_i z^*_{ik} z_{ik}}\langle 0| \ee^{\sum_i z_{ik}^* a_i} 
{\rm e}^{\mathbb{W}\delta t} \ee^{\sum_i z_{i,k-1} a_i^\dag}|0\rangle
\end{equation}
Then, for small enough $\delta t$, one may use the explicit form of $\WW$ to approximate the product as
\begin{equation}
\fl
\ee^{-S_0} = \prod_{k=1}^M \exp\left[ - \sum_i z_{ik}^*(z_{ik}-z_{i,k-1}) - D l_0^{-2} \delta t \sum_{\langle ij\rangle} (z^*_{ik}-z^*_{jk})(z_{i,k-1}-z_{j,k-1}) + O(\delta t)^2 \right]
\label{equ:pi2}
\end{equation}

The presence of finite differences and lattice derivatives makes this expression very unwieldy.
It is therefore conventional to combine (\ref{equ:pi1}) and (\ref{equ:pi2}) into a formal expression, based on the assumption that the time
step $\delta t$ and the lattice space $l_0$ are small enough that all quantities of interest vary little between adjacent points in the space-time
discretisation.  Returning to a general case where $O$ may be any functional of the density, the result is
\begin{equation}
\fl
\langle O \rangle = \int{\cal D}(z,z^*) O[zz^*] \exp\left[ -\int\mathrm{d}t\int\mathrm{d}\bm{r}\, (z^* \partial_t z) - D (\nabla z^*)\cdot(\nabla z) \right]
\label{equ:pi3}
\end{equation}
where the integral over values of $z$ at a large number of space-time points has been replaced by an integral over a (rescaled) function $z(\bm{r},t)$,
and differences between adjacent points have been approximated by considering derivatives of this function.  
the dependence of $\hat{O}$ on $\hat\rho$ is reduced to a simple functional dependence on $z z^*$.
There are boundary
conditions on the functional integral that encode the initial condition $P(0)$ and the behaviour at the final time $t$.  

Finally, to arrive at (\ref{equ:O-pathint}), one makes the change of variables 
\begin{equation}
z z^* = \rho, \qquad z^* = \ee^{\hat\rho}.
\label{equ:zzrho}
\end{equation}
 The transformation has
unit Jacobian so the integration measure at a single spacetime point $\mathrm{d}(z_{ik},z_{ik}^*)$ 
becomes simply $\mathrm{d}\rho_{ik}\mathrm{d}\hat\rho_{ik}$, with the
integration contour for $\hat\rho$ lying along the `imaginary direction' in the complex plane (see also~\cite{itakura2010}).  Within the path integral (\ref{equ:O-pathint}), 
the measure
${\cal D}\rho{\cal D}\hat\rho$ indicates $\prod_{ik} \mathrm{d}\rho_{ik}\mathrm{d}\hat\rho_{ik}$, similarly to the measure in (\ref{equ:pi3}).
As long as the functions $\rho,\hat\rho$
can be safely assumed to be smooth, the transformation (\ref{equ:zzrho}) maps $z^* \partial_t z$ to $\hat\rho \partial_t\rho$ and
$(\nabla z^*)\cdot(\nabla z)$ to $(\nabla\hat\rho)\cdot(\nabla\rho) - \rho|\nabla\hat\rho|^2$.  In the hydrodynamic limit, one expects that that $\rho,\hat\rho$ are indeed smooth enough that higher-order derivatives can be neglected. Thus one makes the passage from
(\ref{equ:pi3}) to (\ref{equ:O-pathint}), which should be valid for large $N$.

\section{Metric structure}
\label{app:metric}

In this section, we sketch the metric structure associated with the Wasserstein geometry, to provide extra context for our main discussion.  On a formal level, the relationships that we quote are straightforward generalisations of results from differential geometry to the space of density profiles, included here for completeness and to illustrate how the geometry of the space of density profiles can be related to the geometry of more familiar finite-dimensional curved spaces. A mathematically rigorous formulation exists but is far beyond the scope of this review~\cite{ags}. 

The main object of interest is the metric tensor.  In physics, metric tensors are familiar from differential geometry: we consider two paths $x(s)$ and $y(s)$ in a $d$-dimensional curved space, with $y(0)=x(0)=a$.  Let the ``tangent vectors'' to these paths at the point $a$ be $u=dx/ds$ and $v=dy/ds$.  Then, the inner product between these two tangent vectors is $u^\mu g_{\mu\nu} v^\nu$, where $g_{\mu\nu}$ is the metric tensor (evaluated at $a$); we use the Einstein convention of implicit summation of repeated upper and lower indices. One can think of $g$ as a symmetric matrix with strictly positive eigenvalues, which ensures that $u^\mu g_{\mu\nu} u^\nu\geq0$, with equality only when $u=0$, as expected for an inner product.
In this curved $d$-dimensional space, the length of a path $x(s)$ is $\ell_d[x(s)] = \int\mathrm{d}s \sqrt{\frac{\mathrm{d}x^\mu}{\mathrm{d}s} g_{\mu\nu}   \frac{\mathrm{d}x^\nu}{\mathrm{d}s}}$ and one may also define a path action ${\cal E}_d[x(s)] = \int\mathrm{d}s\, \frac{\mathrm{d}x^\mu}{\mathrm{d}s} g_{\mu\nu} \frac{\mathrm{d}x^\nu}{\mathrm{d}s}$.  If one minimises, for fixed initial and final position, the functionals ${\cal E}_d$ and $\ell_d$, then one finds that minimisers (critical points) of ${\cal E}_d$ are minimisers (critical points) of $\ell_d$, but the converse is only true is the curve is reparametrised proportional to arc-length (the functional $\ell_d$ is invariant under reparametrisations, ${\cal   E}_d$ is not).  For a path from $s=a$ to $s=b$ one also has
\begin{equation}
\ell_d^2[x(s)] \leq (b-a) {\cal E}_d[x(s)]  .
\label{equ:dist-action}
\end{equation}

To generalise these results to the present context, the $d$-dimensional vector $x$ is replaced by a density profile $\rho=\rho(\bm{r})$.  One may consider the metric tensor $g_{\mu\nu}$ as a bilinear function $g(u,v)=u^\mu g_{\mu\nu} v^\nu$.  The analogue of this function for the Wasserstein geometry in the space of density profiles is
\begin{equation}
G[u_1,u_2] = \int\mathrm{d}\bm{r} \rho \nabla\Phi_1\cdot\nabla\Phi_2
\end{equation}
where the functions $\Phi_{1,2}(\bm{r})$ solve $u_{1,2} = \nabla\cdot(\rho\nabla\Phi_{1,2})$.  Just like $g(u,v)$, the functional $G$ is bilinear, $G(au_1,bu_2) = ab\cdot G(u_1,u_2)$, as well as symmetric under interchange of $u_1$ and $u_2$.  From (\ref{equ:path-cost}) and the associated discussion, one identifies the path cost functional, ${\cal E}[\rho(s)]$ with $\int\mathrm{d}s \, G[\partial_s \rho,\partial_s \rho]$, the natural generalisation of the finite dimensional path action to the Wasserstein geometry.  The result (\ref{equ:d-path-cost}) is therefore analogous to (\ref{equ:dist-action}).
Also, steepest descent processes in finite-dimensional curved spaces obey $g_{\mu\nu}(\mathrm{d}/\mathrm{d}t)x^\nu = -\partial_\mu E$ where $E$ is the energy and $\partial_\mu\equiv \frac{\partial}{\partial   x^\mu}$ as usual.  If $g^{\mu\nu}$ is the inverse metric tensor then one may write instead $(\mathrm{d}/\mathrm{d}t)x^\mu = -g^{\mu\nu}\partial_\nu E$.  Comparing with the discussion of Sec.~\ref{subsec:descent}, we identify the operator ${\rm   grad}^W(\cdot)=-\nabla\cdot\left[\rho\nabla\frac{\delta}{\delta     \rho}(\cdot)\right]$ as the analogue of $g^{\mu\nu}\partial_\nu(\cdot)$.

So far we have shown how various quantities from the main text have analogies in finite-dimensional curved spaces.  Finally, we show how these analogies provide an explanation for the apparently coincidental simplification of (\ref{equ:cost-gradflow}) into the form given in (\ref{equ:cost-gradflow2}).  The path cost (\ref{equ:cost-gradflow}) associated with steepest descent may be identified as
\begin{equation}
{\cal S}[\rho] = \frac{1}{2D} \int_a^b\mathrm{d}t\, G[\partial_t \rho + D \mathrm{grad}^W (\beta F),\partial_t \rho +  D \mathrm{grad}^W(\beta F)].
\label{equ:S1-app}
\end{equation}
However, since the ${\rm grad}^W$ operator is analogous to $g^{\mu\nu}\partial_\nu$, the finite dimensional analogue of this cost is ${\cal S}_d = \frac12 \int\mathrm{d}t\, (\partial_t x^\mu + g^{\mu\nu}\partial_\nu E) g_{\mu\lambda} (\partial_t x^\lambda + g^{\lambda\xi}\partial_\xi E)$, where we use an energy function $E$ as the analogue of $\beta F$, as above.  If we then note that $g_{\mu\lambda}g^{\lambda\xi} =\delta^\xi_\lambda$ since $g^{\mu\nu}$ is (by definition) the inverse metric tensor, then one arrives at 
${\cal S}_d = \frac12 \int\mathrm{d}t [ (\partial_t x^\mu)g_{\mu\lambda} (\partial_t x^\lambda) + 2(\partial_t x^\mu) (\partial_\mu E) +(\partial_\mu E) g^{\mu\nu}(\partial_\nu E)]$.
Noting that $(\partial_t x^\mu) (\partial_\mu E) = \partial_t E$ by the chain rule, we may integrate the second term may to obtain $E(b)-E(a)$, so that
\begin{equation}
\fl
{\cal S}_d = E(b) - E(a) + \frac12 \int\!\mathrm{d}t\, [ (\partial_t x^\mu)g_{\mu\lambda} (\partial_t x^\lambda)  +(\partial_\mu E) g^{\mu\nu}(\partial_\nu E)]. 
\label{equ:Sd2-app}
\end{equation}
Since the last two terms in this equation are time-reversal symmetric while the combination of the first two terms is odd under time-reversal, it follows that if the probability of a path $x(t)$ is given by $\ee^{-{\cal S}_d}$ then there is a detailed balance symmetry between $x(t)$ and its time-reversed
counterpart $x_{\rm TR}(t)$, which is $\mathrm{Prob}[x(t)]\ee^{-E(a)} = \mathrm{Prob}[x_{\rm TR}(t)] \ee^{-E(b)}$.
 To re-cast (\ref{equ:Sd2-app}) in the Wasserstein setting, it is useful to define a bilinear function $\tilde{G}[x,y]=\int\mathrm{d}t \,\rho(\nabla x)\cdot(\nabla y)$ which is the analogue of $x_\mu g^{\mu\nu} y_\nu$.  Then, using the analogy with the finite-dimensional case, in the Wasserstein setting,
\begin{equation}
\fl
{\cal S}[\rho] = \beta F(b) - \beta F(a) + \frac{1}{2D}\int\mathrm{d}t\, G[\partial_t \rho,\partial_t \rho] + 
\frac{D}{2} \int\mathrm{d}t\, \tilde{G}\left[\beta\frac{\delta F}{\delta \rho},\beta\frac{\delta F}{\delta \rho}\right] .
\label{equ:S2-app}
\end{equation}
This result is equivalent to (\ref{equ:cost-gradflow2}) which confirms its validity (it was derived independently in Section~\ref{subsec:descent}).  However, writing this result as in (\ref{equ:S2-app}) emphasises that it is connected to the metric tensor and its inverse, and clarifies its relation to (\ref{equ:S1-app}) and hence to~(\ref{equ:cost-gradflow}).

\end{appendix}


\section*{References}

\begin{thebibliography}{99}

\bibitem{spohn83}
H. Spohn, J. Phys. A {\bf 16}, 4275 (1983)

\bibitem{kov89}
C. Kipnis, S. Olla and S.~R.~S.~Varadhan, Commun. Pure Appl. Math. {\bf 42}, 115 (1989). 

\bibitem{eyink1}
G.~L.~Eyink, J. Stat. Phys. {\bf 61}, 533 (1990).

\bibitem{bertini0102}
L. Bertini, A. De Sole, D. Gabrielli, G. Jona-Lasinio, and C. Landim,
Phys. Rev. Lett. {\bf 87}, 040501 (2001);
J. Stat. Phys. {\bf 107}, 625 (2002).

\bibitem{bertini09}
L.~Bertini, A.~De Sole, D.~Gabrielli, G.~Jona-Lasinio
and C.~Landim, J. Stat. Phys. {\bf 135}, 857 (2009).

\bibitem{bertini14}
L.~Bertini, A.~De Sole, D.~Gabrielli, G.~Jona-Lasinio
and C.~Landim, arXiv:1404.6466.

\bibitem{TKL-prl07}
J. Tailleur, J. Kurchan and V. Lecomte, Phys. Rev. Lett. {\bf 99}, 150602 (2007).

\bibitem{TKL}
J. Tailleur, J. Kurchan and V. Lecomte, J. Phys. A {\bf 41}, 505001 (2008)

\bibitem{imparato09}
A. Imparato, V. Lecomte and F. van Wijland, Phys. Rev. E { \bf 80}, 011131 (2009)

\bibitem{lecomte10}
V. Lecomte, A. Imparato, and F. van Wijland, Prog. Theor. Phys. Supp. {\bf 184}, 276 (2010).


\bibitem{jko}
R.~Jordan, D.~Kinderlehrer, F.~Otto, SIAM J. Math. Anal. {\bf 29}, 1 (1998)

\bibitem{otto2001}
F. Otto, Commun. Partial Differ. Equations {\bf26}, 101 (2001).

\bibitem{ags}
L.~Ambrosio, N.~Gigli, G.~Savar\'e, \emph{Gradient Flows in Metric Spaces and in the Space of Probability Measures (Lectures in Mathematics ETH Z\"urich)}, (Birkh\"auser, 2005)

\bibitem{adams2011}
S. Adams, N. Dirr, M. A. Peletier and J. Zimmer, Commun. Math. Phys. {\bf 307}, 791 (2011).

\bibitem{adams2013}
S. Adams, N. Dirr, M.~A.~Peletier and J. Zimmer, Phil. Trans. Roy. Soc. A {\bf 371}, 20120341 (2013).

\bibitem{touchette-review}
H. Touchette, Phys. Rep. {\bf 478}, 1 (2009).

\bibitem{villani-book}
C.~Villani, \emph{Topics in Optimal Transportation} (Am. Math. Soc., 2003).

\bibitem{benamou}
J.-D. Benamou, Y. Brenier, Numerische Mathematik {\bf 84}, 375 (2000).

\bibitem{doi-peliti}
M.~Doi, J. Phys. A {\bf 9}, 1465 (1976);
L. Peliti, J. Physique {\bf 46}, 1469 (1985).

\bibitem{msr}
P.~C.~Martin, E.~D.~Siggia and H.~A.~Rose, Phys. Rev. A {\bf 8}, 423 (1973).

\bibitem{dedom}
C. De Dominicis, Lett. Nuovo Cimento, {\bf 12}, 567 (1975).

\bibitem{janssen}
H.-K.~Janssen, Z. Phys. B {\bf 23}, 377 (1976).

\bibitem{crooks}
G.~E.~Crooks, Phys. Rev. E {\bf 61}, 2361 (2000).

\bibitem{max}
M.~v.~Renesse, K.-T.~Sturm, {Ann. Probab.} {\bf 37}, 1114 (2009).

\bibitem{hohenberg-halperin}
P. C. Hohenberg, B. I. Halperin, Rev. Mod. Phys. {\bf 49}, 435 (1977).

\bibitem{kmp}
C. Kipnis, C. Marchioro, E. Presutti, 
J. Stat. Phys. {\bf 27}, 65 (1982).

\bibitem{bertini-kmp05}
L. Bertini, D. Gabrielli J.~L.~Lebowitz, J. Stat. Phys. {\bf 121}, 843 (2005).

\bibitem{mouhot}
T. Bodineau, J.~L.~Lebowitz, C.~Mouhot, C.~Villani,  arXiv:1305.7405.

\bibitem{whitelam-metric04}
S.~Whitelam, J.~P.~Garrahan, J. Phys. Chem. B {\bf 108}, 6611 (2004).

\bibitem{ritort-sollich}
F.~Ritort, P.~Sollich, Adv. Phys. {\bf 52}, 219 (2003).

\bibitem{jacquin12-arxiv}
H. Jacquin and F. van Wijland, arXiv:1206.1586.

\bibitem{eyink2}
G.~L.~Eyink, J.~L.~Lebowitz, H.~Spohn, J. Stat. Phys. {\bf 83}, 385 (1996).
  
\bibitem{itakura2010}
K. Itakura, J. Ohkubo and S. Sasa, J. Phys. A {\bf 43}, 125001 (2010)  

\end{thebibliography}
\end{document}